\documentclass[useAMS,usenatbib]{mn2e} 

\usepackage{times,graphicx} 
\usepackage{booktabs} 
\usepackage{amsmath}     
\usepackage{amssymb}     
\usepackage{multicol} 
\usepackage{bm}          
\usepackage{natbib} 
\usepackage[flushleft]{threeparttable} 
\usepackage{calc} 
\usepackage{rotating} 
\usepackage{lscape} 
\bibpunct{(}{)}{;}{a}{}{,} 
\usepackage{longtable} 
\usepackage{enumitem} 
\usepackage{hyperref} 
\usepackage{pifont} 
\usepackage{tikz} 
\usepackage{mathptmx} 

\input{epsf}

\title[NGC\,650-1, a butterfly PN flying in the ISM]{
 The interaction of the halo around the butterfly planetary nebula 
 NGC\,650-1 with the interstellar medium 
}

\author[Ramos-Larios, et al.] 
{G. Ramos-Larios$^{1}$\thanks{E-mail:gerardo@astro.iam.udg.mx (GRL)}, 
M.A. Guerrero$^{2}$, A. Nigoche-Netro$^{1}$, L. Olgu\'{\i}n$^{3}$, M.A. G\'omez-Mu\~noz$^{4,5}$, \newauthor 
L. Sabin$^{6}$, R. V\'azquez$^{6}$, S. Akras$^{7}$, J.C. Ram\'{\i}rez V\'elez$^{6}$ \& M. Ch\'avez$^{8}$
\\ 
$^{1}$Instituto de Astronom\'{\i}a y Meteorolog\'{\i}a, Dpto. F\'{\i}sica,
CUCEI, Universidad de Guadalajara, Av. Vallarta No. 2602, CP 44130,
Guadalajara, Jalisco, Mexico
\\ 
$^{2}$Instituto de Astrof\'{\i}sica de Andaluc\'{\i}a, IAA-CSIC, 
C/Glorieta de la Astronom\'{\i}a s/n, 18008 Granada, Spain
\\ 
$^{3}$Departamento de Investigaci\'on en F\'{\i}sica, Universidad de Sonora, Blvd.\
Rosales-Colosio, Ed. 3H, 83190, Hermosillo, Sonora, Mexico
\\ 
$^{4}$Instituto de Astrof\'{\i}sica de Canarias,  
E-38200, La Laguna, Tenerife, Spain
\\ 
$^{5}$Departamento de Astrof\'{\i}sica, Universidad de La Laguna, 
E-38206, La Laguna, Tenerife, Spain
\\
$^{6}$Instituto de Astronom\'{\i}a, Universidad Nacional Aut\'onoma 
de M\'exico, Apdo. Postal 877, 22860, Ensenada, B. C., Mexico
\\ 
$^{7}$Observat\'orio Nacional/MCTI, 
Rua Gen. Jos\'e Cristino, 77, 20921-400, Rio de Janeiro, Brazil
\\
$^{8}$Instituto Nacional de Astrof\'{\i}sica \'Optica y Electr\'onica, 
Luis Enrique Erro No. 1, 72840, Tonantzintla, Puebla, Mexico
} 

\begin{document} 

\date{Received 2017 July 30th; in original form April 10th} 

\pagerange{\pageref{firstpage}--\pageref{lastpage}} \pubyear{2017} 

\maketitle 

\label{firstpage} 

\begin{abstract} 

With its bright and wide equatorial waist seen almost edge-on 
(\emph{``the butterfly body''}) and the faint and broad bipolar extensions (\emph{``the
butterfly wings''}), NGC\,650-1 is the archetypical example of bipolar
planetary nebula (PN) with butterfly morphology.
We present here deep high-resolution broad- and narrow-band optical images 
that expose the rich and intricate fine-structure of this bipolar PN, with
small-scale bubble-like features and collimated outflows.  
A {\sc SHAPE} spatio-kinematical model indicates that NGC\,650-1
has a broad central torus with an inclination angle of 75$\degr$
with respect to the line of sight, whereas that of the bipolar
lobes, which are clearly seen in the position-velocity maps, is
85$\degr$.
Large field of view deep images show, for first
time, an arc-like diffuse envelope in low- and high-excitation emission
lines located up to 180\arcsec\ towards the East-Southeast of the
central star, well outside the main nebula.
This morphological component is confirmed by \emph{Spitzer} MIPS
and \emph{WISE} infrared imaging, as well as by long-slit low-
and high-dispersion optical spectroscopic observations.
\emph{HST} images of NGC\,650-1 obtained at two different epochs $\sim$14
yrs apart reveal the proper motion of the central star along this direction.
We propose that this motion of the star through
the interstellar medium compresses the remnant material of a slow
Asymptotic Giant Branch wind, producing this bow-shock-like feature.

\end{abstract} 


\section{Introduction}

The morphology of planetary nebulae (PNe) is one of the most important issues 
in their study, displaying an enormous variety of shapes, sizes and complex 
structures. 
The Interacting Stellar Winds (ISW) \citep{Kwok1978} and the Generalized
Interacting Stellar Winds (GISW) models \citep{Balick1987} have made
possible to explain the formation of the most general morphological
classes of PNe by the interaction of the fast and tenuous stellar
wind with the slow and dense Asymptotic Giant Branch (AGB) wind.  
The material ejected during the last phases of the AGB expands and
disperses into the interstellar medium (ISM). 
It can still be seen as very weak and elusive haloes principally of
round shape, but showing fanciful structures in occasions.  
Many authors have done great work concerning the search for halos in PNe, 
demonstrating in a certain way that halos are common in these objects
\citep{Chu1987, Balick1992, Corradi2003}.

Just recently, many serious amateur astronomers have obtained very deep
images of PNe using modest aperture telescopes \citep{Gay2011,Goldman2013}.  
By pushing the sensitivity of the observations using high-efficiency
CCD cameras, dedicated narrow-band filters, and long exposure times,
very low-surface brightness features not detected previously have
been registered, resulting in findings of new structures.

This could be the case of NGC\,650-1
($\alpha$=01$^{\rm h}$42$^{\rm h}$19.6$^{\rm s}$,
$\delta$=+51$\degr$34$^\prime$31\farcm7, J2000),
a.k.a.\ the Little Dumbbell Nebula or M\,76.  
Discovered in 1780 by P.\ M\'echain and recorded the same year by C.\ Messier,
it was W.\ Herschel in 1787 who noticed that the nebula had two lobes barely
separated, leading to the assignment of the two NGC numbers by J.\
Dreyer.
The morphology of NGC\,650-1 is reminiscent of a butterfly shape pattern, with a thick
equatorial waist observed almost edge-on (\emph{``the butterfly body''})
and two broad bipolar lobes (\emph{``the butterfly wings''}).  
At a distance of only 0.93$\pm$0.26 kpc \citep{Frew2016}, it is one
of the most studied PNe. 
Previous works including optical images of NGC\,650-1 exist
\citep[see, e.g.][]{Balick1992,Corradi1995,Hua1997,Ramos2008},
but their resolution and depth are limited. 
The nebula has been described as a central elliptical
ring seen nearly edge-on with two half-shells \citep{Sabbadin1981}.
Its notorious axial symmetry and an inner rectangle with condensations
was noted later \citep{Recillas1984}, and then confirmed as a
bipolar PN with a bright central ring with inner and outer lobes,
the latter with a low expansion velocity of $\sim$ 5 km s$^{-1}$
\citep{Bryce1996}.

Using Hubble Space Telescope (\emph{HST}) images and the spectral photometric
characteristics of the central star, \citet{kf1998} were able to determine
its temperature and \textit{V} magnitude (T$_{eff}$ = 140,000 K, 17.48 mag.,
respectively) and determine a distance to NGC\,650-1 of 1.2 kpc, similar to
that of \citet{Felli1979}.  
The clumpy inner structure of the central bar, possibly produced by
dust \citep{Ramos2008}, suggests extinction variations.  
Radio emission at 6\,cm is reported in this source, with the emission distributed mainly in the central and brightest regions of the core and lobes \citep{Zij1989,Felli1979}. There is also further evidence for molecular emission. \emph{Spitzer} mid-infrared (MIR) IRAC images 
\citep[which probably contain molecular emission][]{2004ApJS..154..296H}
show that the emission at 8 $\mu$m extends over the shorter bands, 
whilst William Herschel Telescope (WHT) 
H$_2$ narrow-band imaging show knots and filaments of material in the core, mostly condensed at the tips of the major axis of the central ring but fainter and barely visible in the bipolar lobes \citep{Marquez2013}. 
\citet{Akras2017} have recently presented the deepest H$_2$ images of PNe [H$_2$ (1-0) S1 and H$_2$ (2-1) S1] and proved that these small structures are also bright in
low-ionization lines.
In this context, it is interesting to highlight the MIR 8 $\mu$m emission
found beyond the main ionized regions, which is due to Polycyclic Aromatic
Hydrocarbon (PAH) band emission \citep{Ramos2008}.
It should be noted that CO molecular emission was not detected
\citep{Huggins1989}, although NGC\,650-1 is a carbon-rich (C/O$>$2)
nebula \citep[]{Kwitter1996,vanHoof2013}.

Photodetector Array Camera and Spectrometer (PACS) and Spectral and Photometric Imaging 
Receiver (SPIRE) images from the \emph{Herschel} telescope found considerable extinction of ultraviolet  (UV) photons inside the main torus of the nebula \citep{steene2012}. 
In this sense, an excellent work performed by \citet{vanHoof2013} with the same \emph{Herschel} observations and using photoionization models found that the dust grains are large (0.15 $\mu$m) in the ionized main nebula, being heated both by stellar emission and diffuse Ly$\alpha$ emission from the ionized gas.
They also derived a very high temperature of 208,000 K (50\% higher
than previously published values) for the central star.

The interaction between PNe and the surrounding ISM has been subject
of various studies.
Indeed, understanding this process will give us valuable clues about
the local characteristics of the ISM as well as the chemistry, dynamics
and kinematics of PNe.
Hence, several objects showing a PN-ISM interaction have been discovered
and studied \citep{Borkowski1990,Tweedy1996,Xilouris1996,Sabin2010}.
Most have been recently compiled and analyzed by \citet{Ali2012,Ali2013}.
The observations of these PNe clearly show different degrees or stages
of interaction with the ISM, ranging from the sole interaction of the
halo (remnant of a former AGB ejection or thermal pulse) surrounding
the unaffected PN to the totally disrupted PN shell merging with the
ISM.
\citet{Wareing2007} used hydrodynamical modelling and issued a classification
(dubbed ``WZO'') associated to the physical conditions of both the ISM and
the PNe that declines into four stages of interaction WZO1--4.
In particular, their WZO1 phase regroups all the objects in their
early phase of interaction where the only affected component (in
the form of a bow-shock) is the surrounding faint halo of the
otherwise still morphologically intact PN.
The number of PNe in the WZO1 stage represents $\sim$26\% of the whole
sample of 117 objects compiled by \citet{Ali2012}, based mainly on
observations coming from \citet{Ramos2009} and \citet{Corradi2003}.
A cross check with the Hong Kong/AAO/Strasbourg H$\alpha$ PN
database (HASH) \citep{Parker2016} reveals 9 bipolar PNe with
interacting halo:
Cn\,1-5, M\,2-40, NGC\,2440, NGC\,3242, NGC\,6765, NGC\,6772,
NGC\,6804, NGC\,6853, and NGC\,7293.  
NGC\,650-1 might add to this group.

\begin{table}\centering 
\setlength{\columnwidth}{0.1\columnwidth} 
\setlength{\tabcolsep}{1.0\tabcolsep} 
\caption{Optical Imaging} 
\begin{tabular}{lllrr} 
\hline 

\multicolumn{1}{c}{Telescope} & 
\multicolumn{1}{c}{Instrument} & 
\multicolumn{1}{c}{Filter} & 
\multicolumn{1}{c}{$\lambda_{c}$} & 
\multicolumn{1}{c}{$\Delta\lambda$} \\ 

\multicolumn{1}{c}{} & 
\multicolumn{1}{c}{} & 
\multicolumn{1}{c}{} & 
\multicolumn{1}{c}{(\AA)} & 
\multicolumn{1}{c}{(\AA)} \\ 
\hline 
\hline 

IAC80     & CAMELOT & [O~{\sc iii}]  & 5007 & 30~~~~ \\ 
       &         & H$\alpha$      & 6567 & 8~~~~ \\ 
       &         & [N~{\sc ii}]   & 6571 & 50~~~~ \\\\ 
NOT     & ALFOSC  & [O~{\sc iii}]  & 5007 & 8~~~~ \\ 
       &         & H$\alpha$      & 6567 & 8~~~~ \\ 
       &         & [N~{\sc ii}]   & 6588 & 9~~~~ \\\\ 
GTC     & OSIRIS  & g'             & 4815 & 1530~~~~ \\ 
       &         & r'             & 6410 & 1760~~~~ \\ 
       &         & OS657          & 6570 & 350~~~~ \\ 
\hline 
\hline 
\end{tabular} 
\vspace{0.3cm} 
\end{table} 

With the purpose to investigating the real shape of this diffuse halo,
their kinematics, composition and ISM interaction in NGC\,650-1, we
have used \emph{Spitzer} MIPS and \emph{WISE} infrared images, new deep
and high-resolution broad- and narrow-band optical images, and long-slit
low- and high-dispersion spectroscopic observations.
These have been examined in conjunction with the {\sc SHAPE} program
\citep{Steffen2011} in order to establish a simplified model of the
whole PN.

\begin{figure*} 
\centering 
\includegraphics[height=5in]{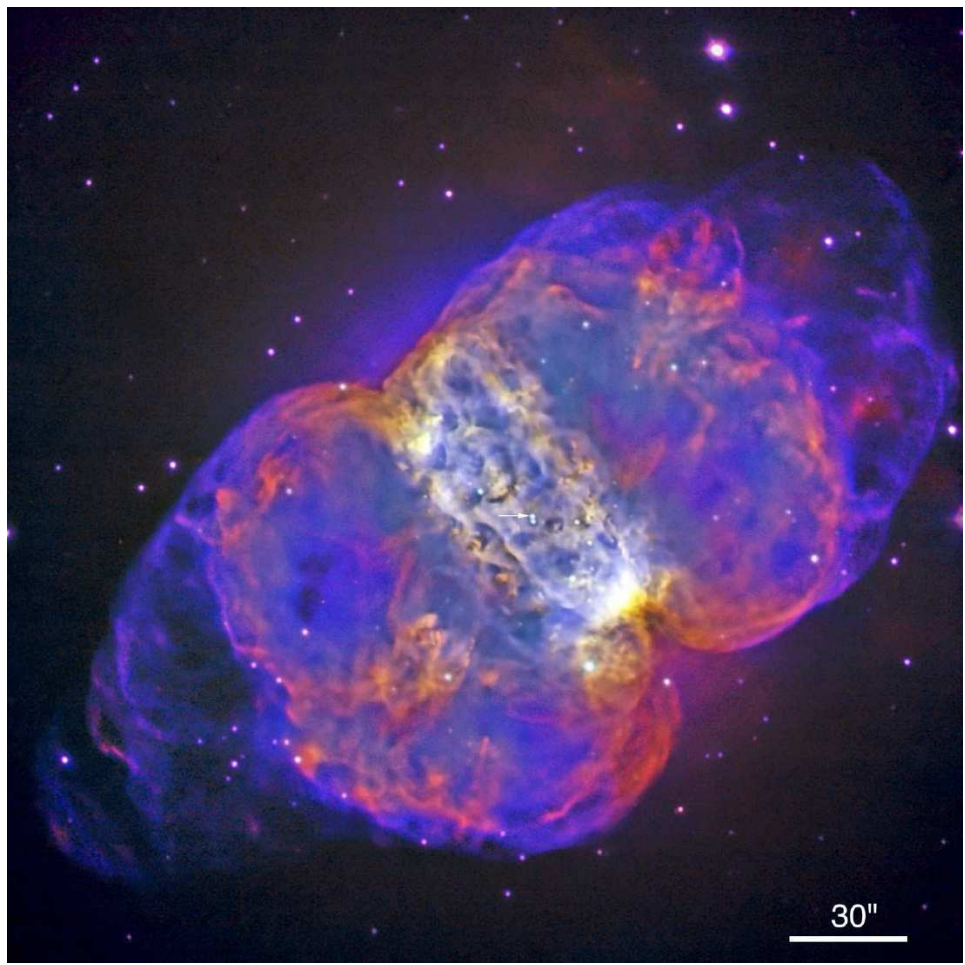} 
\caption{High-resolution colour hybrid composite combined picture of NGC 650-1 obtained through 
[O~{\sc iii}] $\lambda$5007 (blue), H$\alpha$ (green), and [N~{\sc ii}] $\lambda$6583 (red) narrow-band filters at the NOT, and g' (blue), r' (gren)
and OS657 (red) broad-band filters at the GTC (see text for details). 
In this picture, north is up, east to the left.  
This picture highlights the wealth of features in the main nebula 
associated with the bipolar lobes and equatorial ring. 
These show multiple clumps and filaments, mostly in emission, but also 
seen as dark absorptions projected onto the equatorial ring. 
The position of the central star is marked at the centre of the nebula
by a small arrow.
Note the hints of diffuse [O~{\sc iii}] and [N~{\sc ii}] barely visible 
just outside the nebular waist.   
} 
\label{GTC.img} 
\end{figure*} 

The observations and archival data are presented in \S\ref{sec_obs}
and \S\ref{sec_irobs}. 
The results on the morphology are described in \S\ref{sec_mor}.
The spectroscopy is discussed in \S\ref{spec_sec}.
The kinematics, model and the discussion are presented
in \S\ref{sec_dis} and final conclusions are presented
in \S\ref{sec_con}.

\section[]{Optical Observations}\label{sec_obs} 

\subsection{GTC imaging} 

A high-resolution broad-band colour composite image in the g' and r' Sloan filters and 
the Order Sorter (OS) OS657 filter was obtained from the Gran Telescopio Canarias (GTC) 
Astronomical Images Gallery (http://www.gtc.iac.es/multimedia/imageGallery.php).
The series of images were acquired with the imager and spectrograph OSIRIS (Optical 
System for Imaging and low-Intermediate-Resolution Integrated 
Spectroscopy) at the Nasmyth-B focus of the GTC at the 
Observatorio del Roque de Los Muchachos (ORM), La Palma, Spain.   
The detector consists of a mosaic of two 2048$\times$4096 Marconi CCD42-82 with pixel size of 
15$\mu$m and a binned 2$\times$2 plate scale of 0$\farcs$254 arcsec~pix$^{-1}$, resulting in a field 
of view (FoV) of 7$\farcm$8$\times$8$\farcm$5 (7$\farcm$8$\times$7$\farcm$8 unvignetted).
Three 30\,s exposures were obtained in the g' and r' filters, 
and another three 60\,s exposures were taken using the OS657 
filter. 
The central wavelengths and bandwidths of all the optical observations 
filters are listed in Table~1. 

\begin{figure*} 
\centering 
\includegraphics[height=5in]{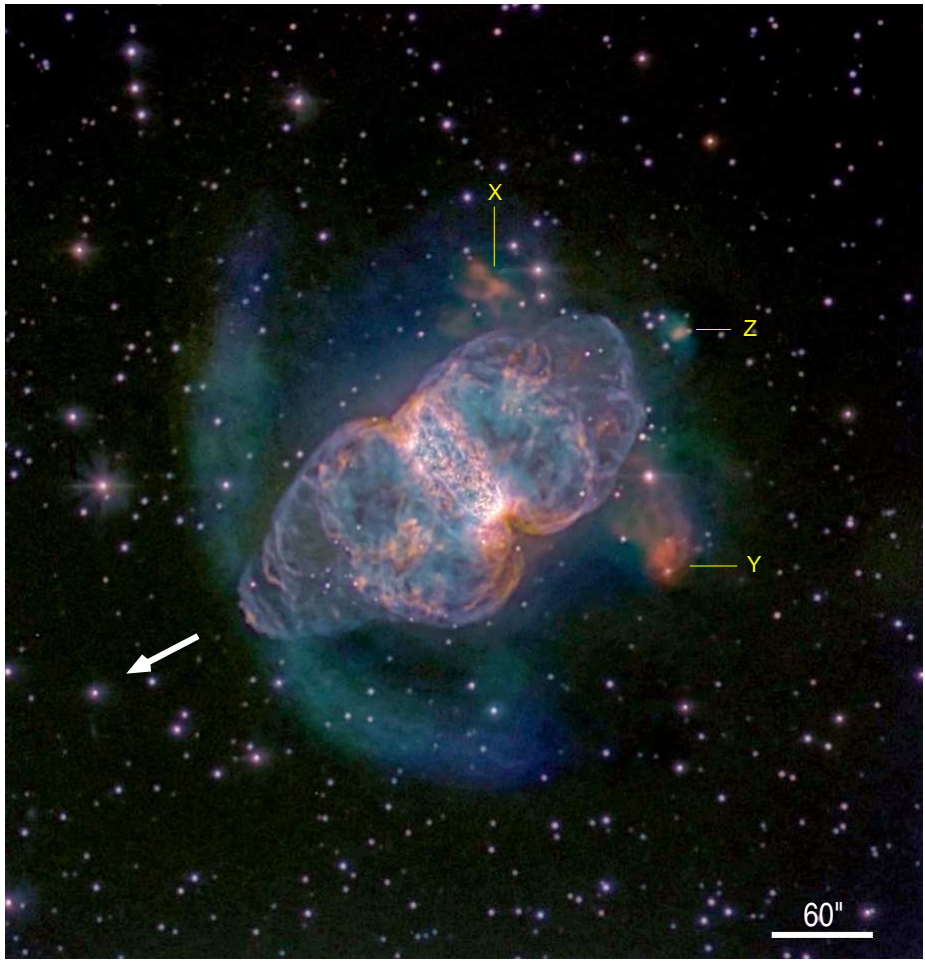} 
\caption{ 
NOT and IAC80 [O~{\sc iii}] $\lambda$5007 (blue), H$\alpha$ (green), 
and [N~{\sc ii}] $\lambda$6583 (red) \textit{RGB} hybrid composite combined picture 
of NGC\,650-1. 
As in the image in Figure~\ref{GTC.img}, north is up, east to the left.
This combination improves the spatial resolution of the main nebula, 
and emphasizes the outer emission. 
Extended emission is clearly visible at distances up to 180\arcsec\ 
from the central star, with an arc-shaped feature of smooth diffuse 
H$\alpha$ and [O~{\sc iii}] (green-blue) emissions pointing towards 
the East-South east, and [N~{\sc ii}]-bright (red) patches and blobs (marked X,Y and Z) 
towards the West of the main nebula. 
The direction of the motion of the central star is marked with the big white arrow.
The [N~{\sc ii}] emission is mostly coincident with \emph{Spitzer} 
MIR emission \citep{Ramos2008}. Credit IAC80 image: Daniel L\'opez.
} 
\label{650.img} 
\end{figure*}

\subsection{NOT imaging} 

High-resolution [O~{\sc iii}], H$\alpha$, and [N~{\sc ii}] narrow-band 
images were obtained on September 3, 2008 using ALFOSC (Andalucia Faint 
Object Spectrograph and Camera) at the 2.56-m Nordic Optical Telescope 
(NOT) of the ORM.   
The detector was a 2048$\times$2048 CCD with plate scale 
0$\farcs$19 arcsec~pix$^{-1}$, resulting in an FoV of 
6$\farcm$5$\times$6$\farcm$5.   
Two individual frames with integration times of 300\,s were taken for 
each filter.   
All the data were bias-subtracted and flat-fielded using twilight flats 
employing standard {\sc IRAF} routines. 
The spatial resolution, as derived from the full width half maximum (FWHM)
of stars in the FoV, 
was $\sim$0$\farcs$65.
A composite picture of the GTC and NOT combined images is shown in
Figure~\ref{GTC.img}.

\subsection{IAC80 imaging} 

A very deep [O~{\sc iii}], H$\alpha$, and [N~{\sc ii}] narrow-band colour composite image 
was downloaded from the Banco de Im\'agenes Astron\'omicas (http://www.bia.iac.es/). 
The images were acquired on 2008 November 29, 2008 using the CAMELOT (CAmara MEjorada 
L\textbf{i}gera del Observatorio del Teide) camera mounted at the 0.82-m IAC80 
(Instituto de Astrof\'{\i}sica de Canarias) Telescope at the Observatorio 
del Teide in Tenerife, Spain.

A 2048$\times$2048 CCD was used as detector.   
Its pixel size of 13.5$\mu$m implies a plate scale of
0\farcs304~pixel$^{-1}$, resulting in an FoV of
10\farcm4$\times$10\farcm4.   
Four frames with integration times of 1800\,s each were taken for 
the H$\alpha$ and [N~{\sc ii}] filters, and six frames with similar 
integration times were taken for the [O~{\sc iii}] filter. 
All the data were processed using {\sc MaxIm DL}\footnote{ 
 Suite of image acquisition, processing, and analysis tools 
 by Cyanogen Imaging. 
} software. 
The seeing, as measured from stars in the FoV, was $\sim$0\farcs9.   
A composite combined picture of the NOT and IAC80 images is shown in
Figure~\ref{650.img}.

\subsection{Long-slit high-dispersion spectroscopy} 

Long-slit high dispersion optical spectroscopy of NGC\,650-1 
was obtained on October 21-22 2014 using the Manchester Echelle 
Spectrometer \citep[MES,][]{Meaburn2003} mounted on the 
2.1\,m (f/7.5) telescope at the Observatorio Astron\'omico 
Nacional de San Pedro M\'artir (OAN-SPM, Mexico).   
A 2k$\times$2k CCD with pixel size of 13.5 $\mu$ pix$^{-1}$ was used
as detector with a 2$\times$2 on-chip binning, resulting on a plate
scale of 0\farcs351\,pix$^{-1}$. 
Since MES has no cross dispersion, a $\Delta\lambda$=90\,\AA\ bandwidth 
filter was used to isolate the 87$^{\rm th}$ order covering the spectral 
range which includes the H$\alpha$ and [N~{\sc ii}] $\lambda6583$ lines,
with a dispersion of 0.05\,{\AA}~pix$^{-1}$.   
The [O~{\sc iii}] $\lambda$5007 ($\Delta\lambda = 50$) filter 
was used to isolate the 114$^{\rm th}$ order, corresponding to 
0.043\,{\AA}~pix$^{-1}$. 
At this spectral orders, the slit width of 150-$\mu$m (2\arcsec) that 
was set during the observations corresponds to a spectral resolution 
of $\simeq12$\,km\,s$^{-1}$.

\begin{figure} 
\begin{center} 
\includegraphics[width=0.8\columnwidth]{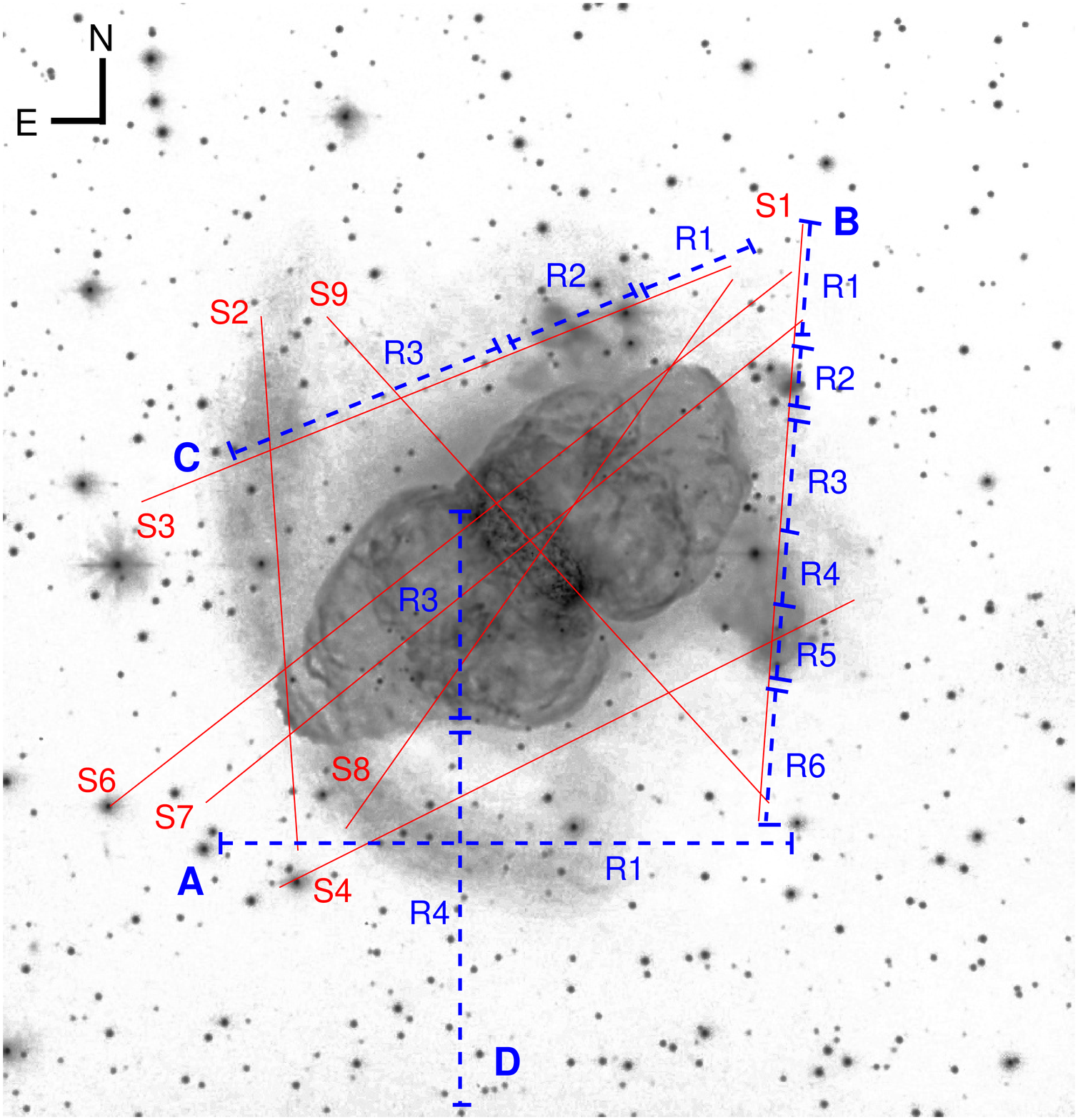} 
\vskip .1in 
\includegraphics[width=0.8\columnwidth]{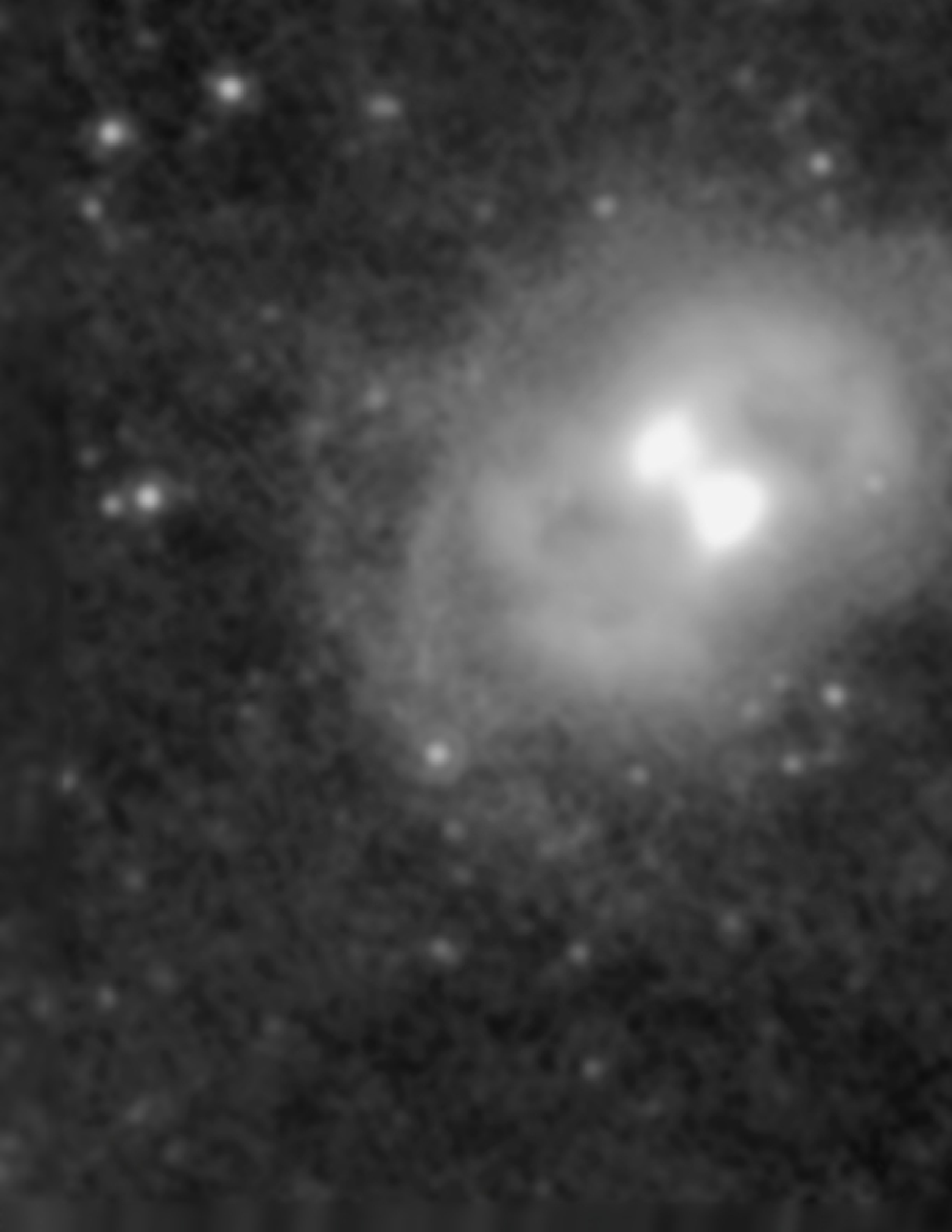} 
\vskip .1in 
\includegraphics[width=0.8\columnwidth]{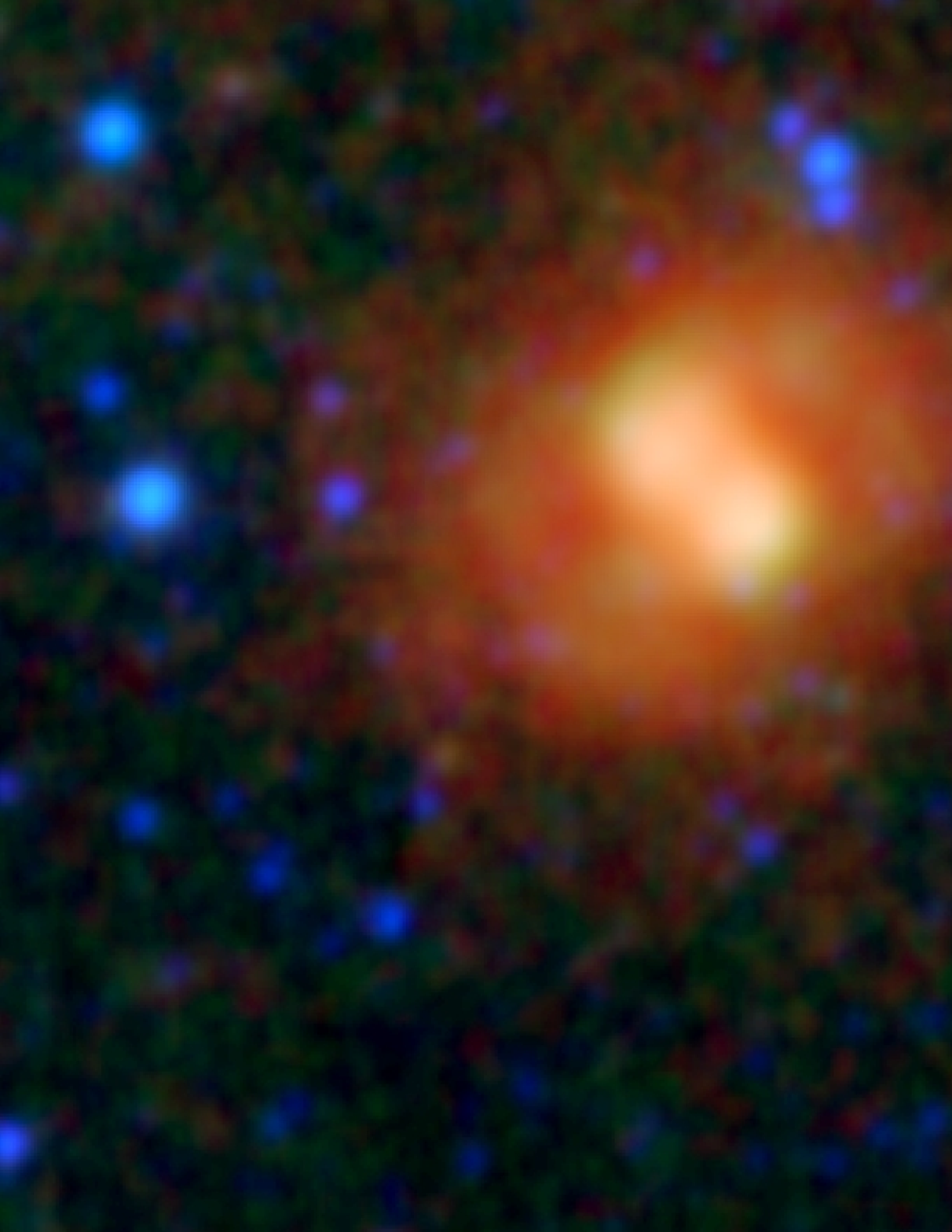} 
\vskip .1in 
\end{center} 
\caption{
NOT-IAC80 {\it (top)}, \emph{Spitzer} MIPS 24 $\mu$m {\it (middle)},
and \emph{WISE} three-band colour composite {\it (bottom)} pictures
of NGC\,650-1.
The NOT-IAC80 image has been labelled with the position of the long-slits
used for the high- (solid red lines) and low-resolution (dashed blue lines)
spectra.
The regions used for the extraction of one-dimensional low-resolution
spectra are marked on the four long-slits marked as A, B, C, and D.  
} 
\label{ir.img} 
\end{figure}

Eight long-slit spectra, marked with the red labels S1 to S4 and S6 to S9 
on the NOT-IAC80 greyscale image in the top panel of Figure~\ref{ir.img}, 
were obtained to map the kinematics at different regions of the nebula. 
The position angles (PAs) for slits S1 to S4, measured in
the conventional way anticlockwise from the north, are $-3\degr$, 
$+3\degr$, $+110\degr$, and $+120\degr$, respectively, whilst
for slits S6 to S9 are $-49\degr$, $-55\degr$, $-36\degr$,
and $37\degr$, respectively. 
The exposure time was 1,800\,s for the spectra S1 to S4 and
1,200\,s for the rest. 
The seeing during the observations, as determined from the 
FWHM of stars in the FoV, varied from 1\farcs2 to 2\farcs0. 
The spectra were wavelength calibrated with a ThAr arc lamp to an accuracy 
of $\pm1$\,km\,s$^{-1}$ using {\sc IRAF}.

\label{kin_model} 
\begin{figure} 
\centering 
\includegraphics[width=0.49\columnwidth]{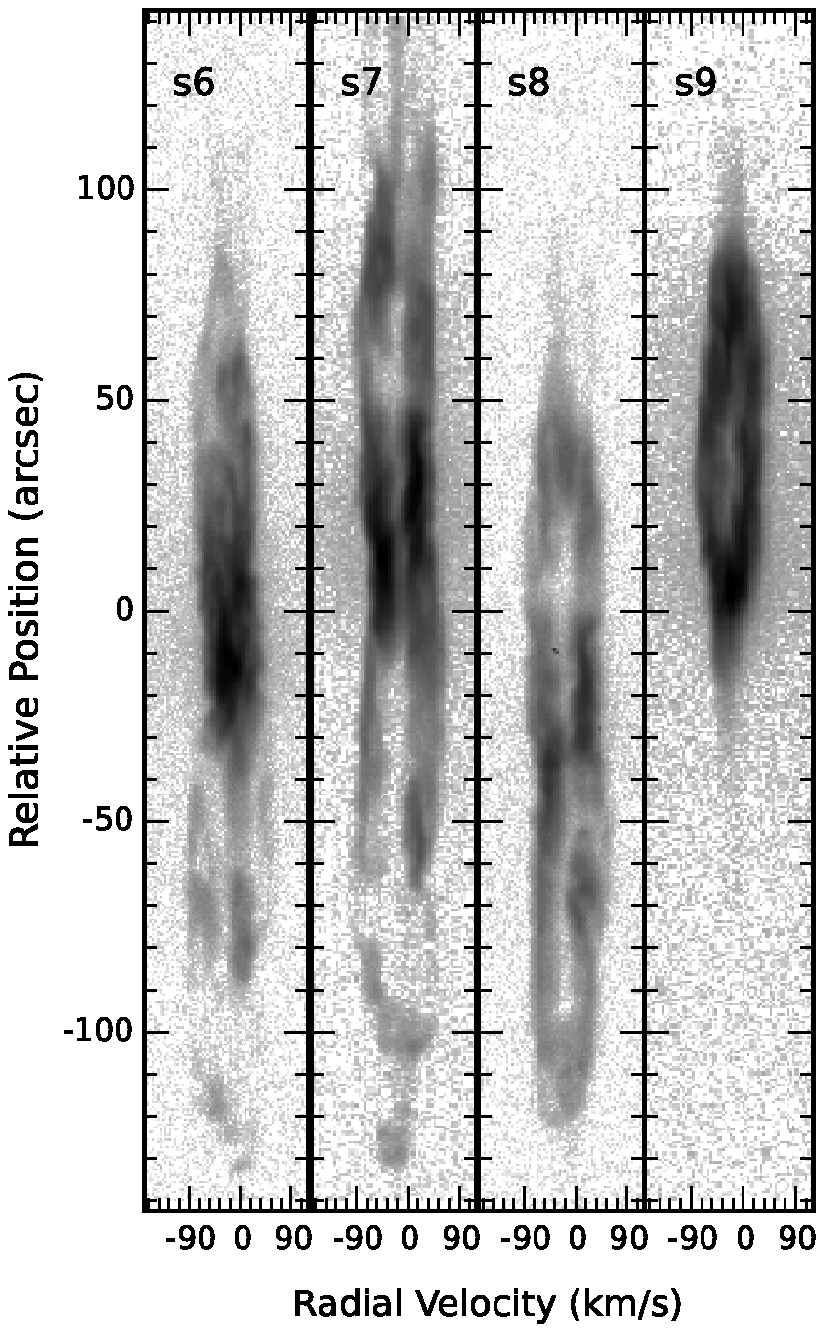} 
\includegraphics[width=0.49\columnwidth]{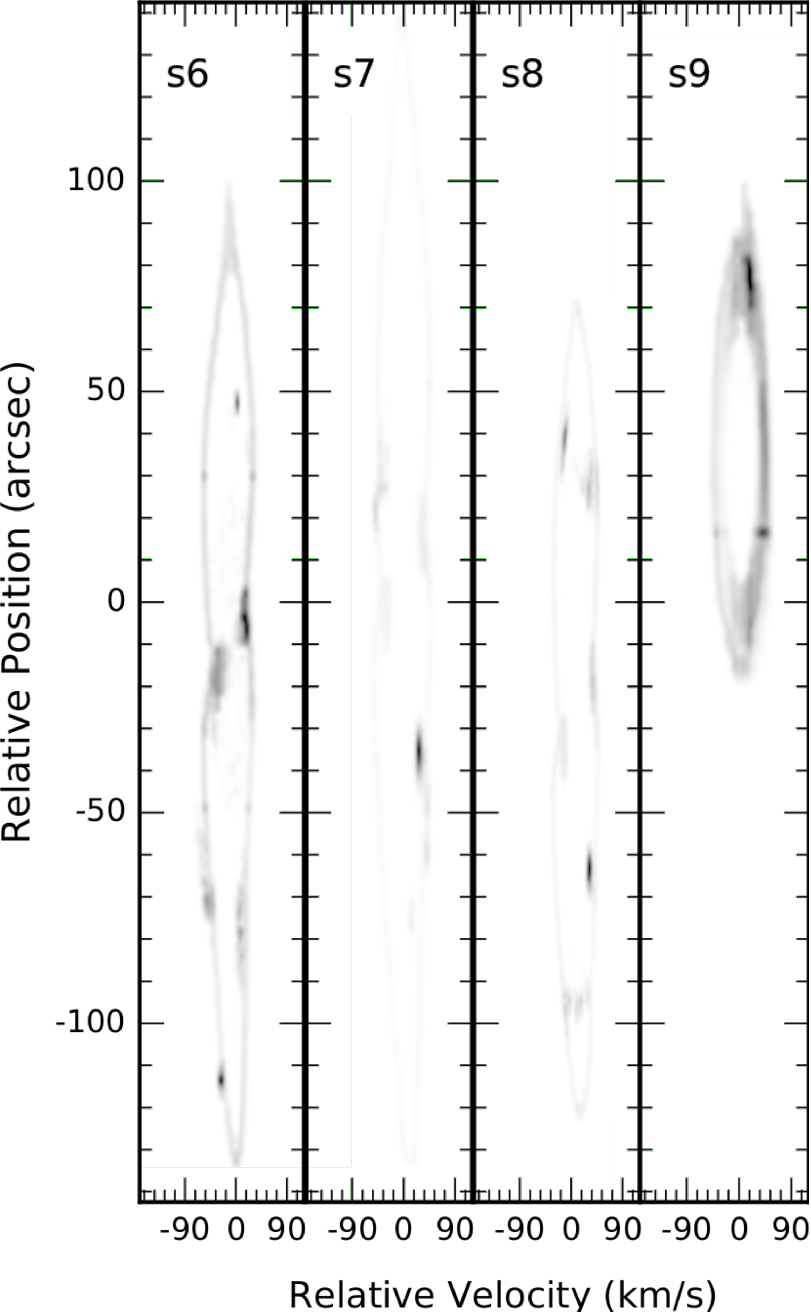} 
 \caption{ 
 NGC\,650-1 MES position-velocity (PV) map in the [O~{\sc iii}] emission
  line ({\it left}) derived from the long-slits covering the main nebula
  (S6, S7, S8, and S9) and synthetic PV maps derived from our {\textsc SHAPE}
  model ({\it right}). 
} 
\label{p1.img} 
\end{figure} 

\subsection{Long-slit low-dispersion spectroscopy} 

A series of low dispersion, long-slit optical spectra of NGC\,650-1 were obtained 
on October 27-28, 2014, using the Boller \& Chivens (B\&C) spectrometer mounted at the 
prime focus of the OAN-SPM 2.1m telescope.   
A 2048$\times$2048 CCD was used as a detector, in conjunction with 
a 400 lines mm$^{-1}$ grating blazed at 5500 \AA.   
The slit had a length of 5\arcmin\ and a width of 200 $\mu$m 
($\equiv$2\arcsec).   
The plate and spectral scales were 0\farcs57~pixel$^{-1}$ and 1.7~\AA~pixel$^{-1}$, 
respectively. The spectral resolution was $\sim$4 \AA, the wavelength uncertainty was $\sim$1 \AA, and the spectral range covered was 4080--7560 \AA. 
Three 1800\,s exposure were obtained for every slit oriented through four main directions of PAs= 90, $-$4, 110, and 0 {\degr} named A, B, C and D respectively (marked in colour blue in the top panel of Figure~\ref{ir.img}).   
The mean air mass during the observations was $\simeq$1.1.   
The observations were flux calibrated using a 300\,s exposure of the star 
Hiltner\,600 obtained on October 28.   
The seeing as determined from the FWHM of stars in the FoV, was 
$\simeq$2\farcs1. 
The spectrum was bias-subtracted, flat-fielded, wavelength (CuAr lamp) and 
flux calibrated following standard procedures using {\sc XVISTA}.

\section[]{Infrared Data}\label{sec_irobs} 

\subsection{WISE imaging}

Wide-field Infrared Survey Explorer Space Telescope 
\citep[\emph{WISE},][]{2010AJ....140.1868W}
observations of NGC\,650-1 obtained in the W2, W3,
and W4 bands were retrieved from the NASA/IPAC
Infrared Science Archive (IRSA). 
\emph{WISE} surveyed the entire sky at 3.4, 4.6, 12, and 22 $\mu$m
(W1 to W4 bands) with angular resolutions of 6\farcs1, 6\farcs4,
6\farcs5, and 12\farcs0, respectively.
The system consists of a 0.4m telescope using HgCdTe and
Si:As 1024$\times$1024 detector arrays with a plate scale
of 2\farcs75~pixel$^{-1}$.     
Its astrometric accuracy for bright sources is better than 0\farcs15. 
The \emph{WISE} RGB composition image (bottom panel of Figure~\ref{ir.img}) where 
the W2 (blue), W3 (green) and W4 (red) bands were used,
resembles the outer shape of the Figure~\ref{650.img}, with the east-side
of the nebulae showing a sharp boundary. 

\begin{figure} 
\centering  
\includegraphics*[width=0.525\linewidth]{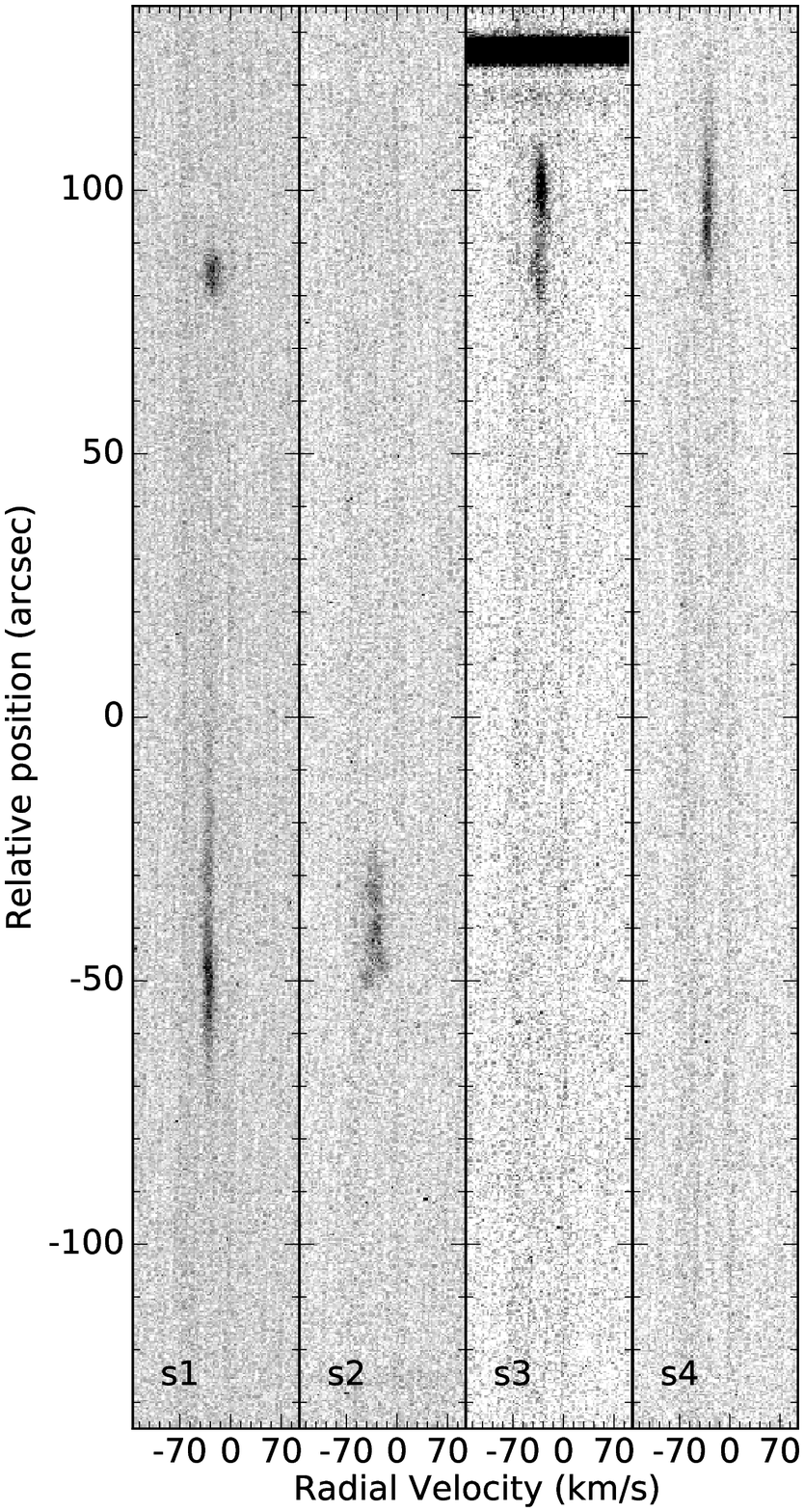} 
\includegraphics*[width=0.45\linewidth]{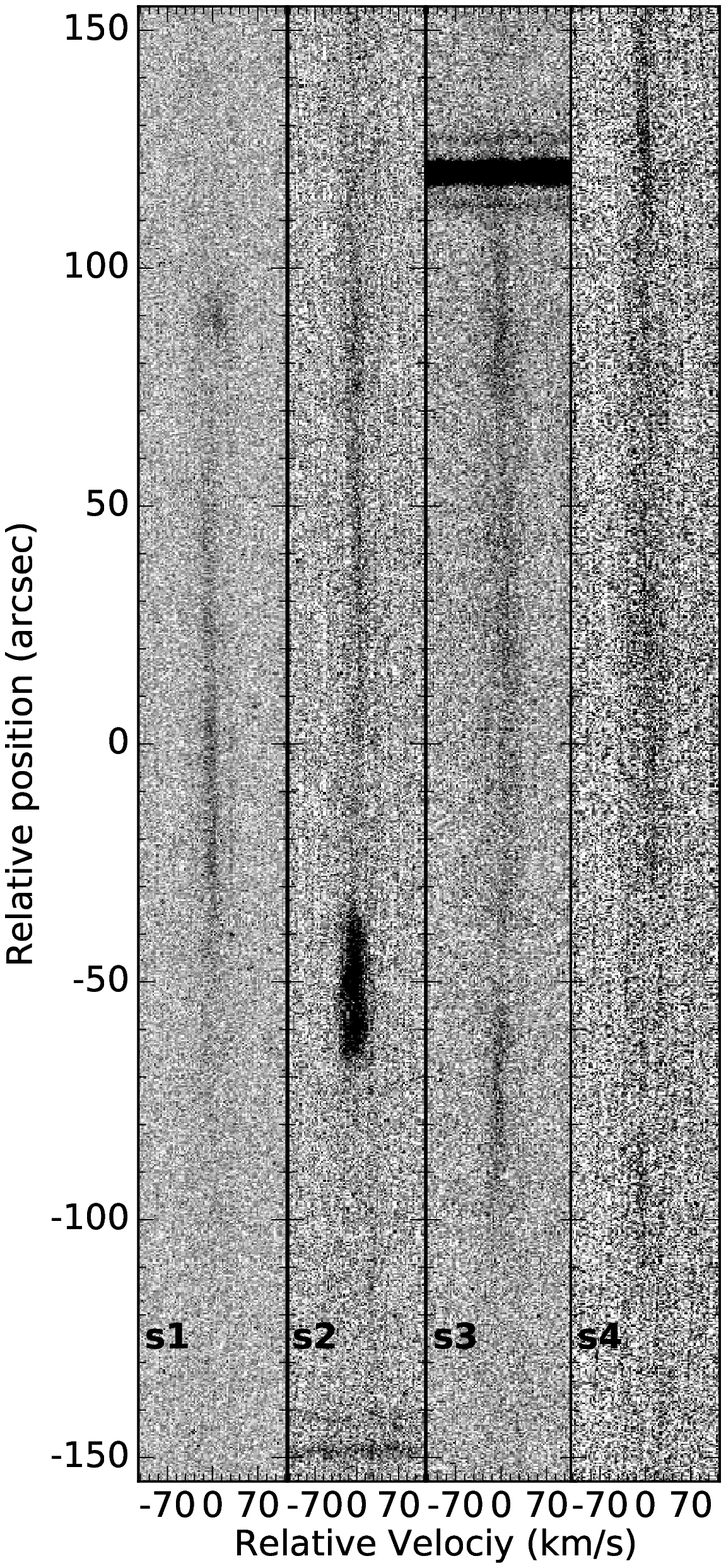} 
\caption{
 NGC\,650-1 MES position-velocity (PV) maps in the [N~{\sc ii}] ({\it left})
  and [O~{\sc iii}] ({\it right}) emission lines derived from the long-slits
  covering the halo (S1, S2, S3, and S4). 
} 
\label{p2.img} 
\end{figure} 

\subsection{Spitzer MIPS images}

\emph{Spitzer} Space telescope Multiband Imaging Photometer for Spitzer
\citep[MIPS,][]{Rieke2004} observations of NGC\,650-1 were retrieved
from the NASA/IPAC Infrared Science Archive (IRSA).
NGC\,650-1 was observed on September 2004 under the programme ID 77
(Stellar Ejecta: Macro-Molecule and Dust Formation and Evolution,
PI: G.\ Rieke). 
MIPS produces image and photometry in three main bands, centred at
24 $\mu$m, 70 $\mu$m, and 160 $\mu$m, and low-resolution spectroscopic
data in the range from 55 to 95 $\mu$m. 
The MIPS detector consists of a Si:As 128$\times$128 pixels array.  
The pixel size of 24 $\mu$m results in an FoV of 5\farcm4$\times$5\farcm4.
The MIPS observations were performed in the scan-mapping mode, used
to image large areas of the sky in one or more bands simultaneously. 
The \emph{Spitzer} 24 $\mu$m image (middle panel of Figure~\ref{ir.img})
shows the main nebula and the sharp edge of the outer structure towards
the south-east.

\section{Morphology}\label{sec_mor}

The optical GTC, NOT, and IAC80 images and the archival IR \emph{Spitzer} 
and \emph{WISE} images have been presented in Figures~\ref{GTC.img},
\ref{650.img}, and \ref{ir.img}. 
The GTC and NOT unsharp masking colour \textbf{hybrid} composite picture in 
Figure~\ref{GTC.img} reveals fine details of the bipolar lobes and 
equatorial ring of NGC\,650-1.  
The hybrid composite picture involves layering the higher resolution GTC 
image onto the lower resolution NOT image.  
In this case, this step enhances the small-scale structure features at the 
core region of NGC\,650-1 seen in the GTC from the GTC image, whereas the 
more extended emission from the bipolar lobes present in the NOT image are 
emphasized.  
Meanwhile, the NOT and IAC80 unsharp masking colour \textbf{hybrid} 
composite picture in Figure~\ref{650.img} brings to life a new complex 
of low surface-brightness outer structures surrounding the main nebula. 
Here, the fainter large-scale structures are best imaged by the IAC80, 
with its larger pixel size and coarser spatial resolution, whereas the 
brighter emission from the main nebula is best resolved by the smaller 
pixel size and finer spatial resolution of the NOT images.  
Finally, the IR \emph{Spitzer} and \emph{WISE} images in Figure~\ref{ir.img} 
complement the optical features with a view of the dusty outer component of 
NGC\,650-1 and ionized main nebula as seen in IR emission lines.
The different morphological components of NGC\,650-1 are described next.

\subsection{Main Nebula}

The optical morphology of the main nebula is best seen in the NOT and 
GTC colour-composite picture presented in Figure~\ref{GTC.img}. 
The main nebula consists of an equatorial ring and two broad bipolar 
lobes with prominent polar protrusions.   
The bipolar morphology of NGC\,650-1, which is very similar to that of
other butterfly-shaped PNe such as Sh\,1-89 \citep{Manchado1996}, has
been described many times in the literature
\citep[e.g. ][]{Balick1987,Sabin2012,Guerrero2013,Uscanga2014,Manchado2015}.   
In this section we will focus on the new details revealed by
the NOT and GTC images.

The equatorial ring has a size $\sim$95\arcsec\ and is oriented 
along PA $\sim$35$^\circ$. 
There is a notable ionization gradient, with the [N~{\sc ii}] emission
distributed mostly along the outer edges of the ring, the [O~{\sc iii}]
emission mostly on the central regions, and the H$\alpha$ emission
delineating the ring at intermediate radial distances from the central
star.  
The ring interior is filled with emission that follows an intricate 
pattern of filaments and patches, but it also reveals dark filaments 
and blobs that absorb the emission in the background. 
These higher density regions may be associated with the molecular-rich
filaments and knots with bright H$_2$ emission \citep{Marquez2013,Manchado2015}.

Diffuse [O~{\sc iii}] and [N~{\sc ii}] emission is faintly visible towards 
the North and South of the nebular waist, respectively.   
A few arc-like features are hinted in the southern region, although 
these smooth features are not as clear as those detected in 
IC\,4406 \citep{Ramos2016} and NGC\,2346 \citep{Phillips2010}, where it has been 
presumed the possible existence of ring-like structures associated with 
the detailed mass-loss in the last phases of the AGB 
\citep{Corradi2004,Kim2017}.

Two broad bipolar lobes protrude from the equatorial ring. 
The bipolar lobes are especially prominent in [N~{\sc ii}], with
emission concentrated along the outer edges of the bipolar lobes
(red in Figures~\ref{GTC.img} and \ref{650.img}).
The [N~{\sc ii}] emission seen inside the bipolar lobes most likely
corresponds to regions of bright [N~{\sc ii}] emission at the walls
of the bipolar lobes projected inside them. 
There are many interesting features, such as radial filaments (in red
in Figure~\ref{GTC.img}) pointing towards the central star and dark
patches indicating the presence of foreground absorbing material. 

The broad butterfly-shaped bipolar lobes show extensions at their tips, 
with a single bow-shock feature towards the Southeast and several (at 
least three) similar structures towards the Northwest. 
The emission from these structures is mostly dominated by [O~{\sc iii}], 
may be revealing a higher excitation caused by illumination effects or
by shocks \citep{Guerrero2000}.

\subsection{Extended Outer Emission} 

The optical morphology of the extended outer emission is best seen in the 
NOT and IAC80 colour-composite picture presented in Figure~\ref{650.img}. 
The outer emission consists of an arc-like feature located towards 
the East-Southeast of the main nebula and [N~{\sc ii}] blobs above 
and below the western bipolar lobe of NGC\,650-1.   

\begin{figure} 
  \begin{center}
    \vspace*{-0.2cm}
\includegraphics[bb=42 16 576 784,width=0.7\linewidth, angle =-90]{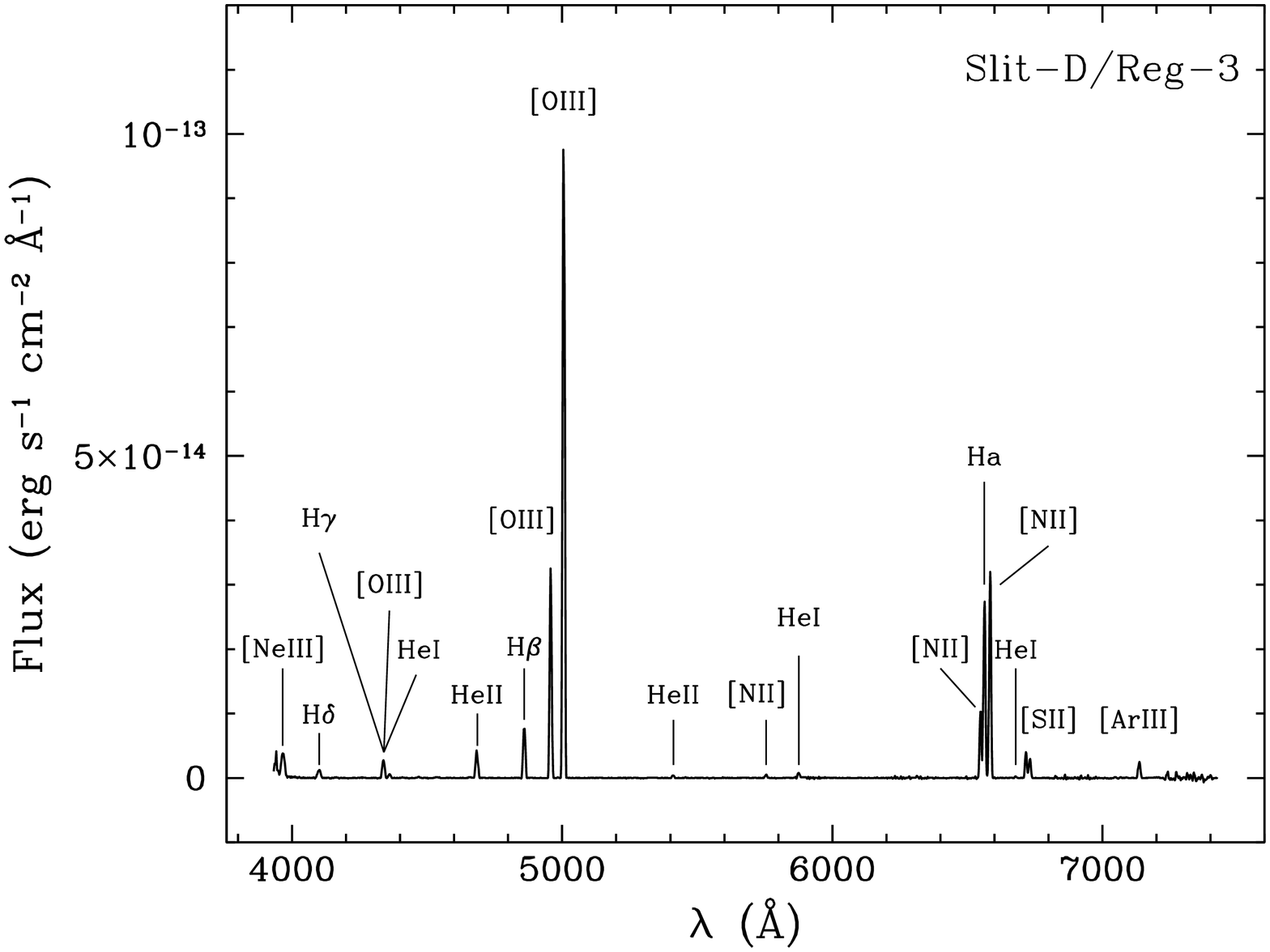} 
\vskip .1in 
\includegraphics[bb=42 16 576 784,width=0.7\linewidth, angle =-90]{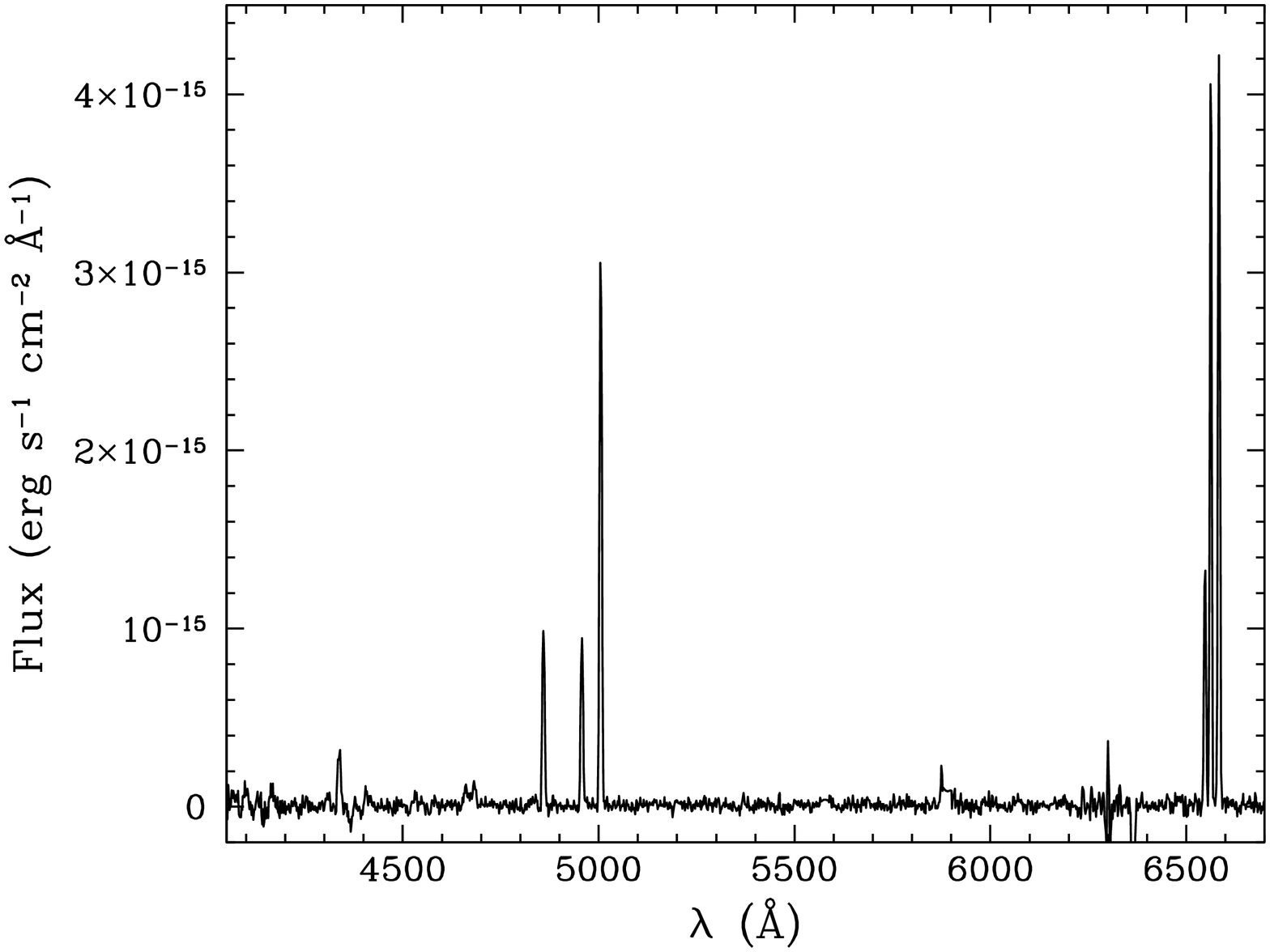} 
\vskip .1in 
\includegraphics[bb=42 16 576 784,width=0.7\linewidth, angle =-90]{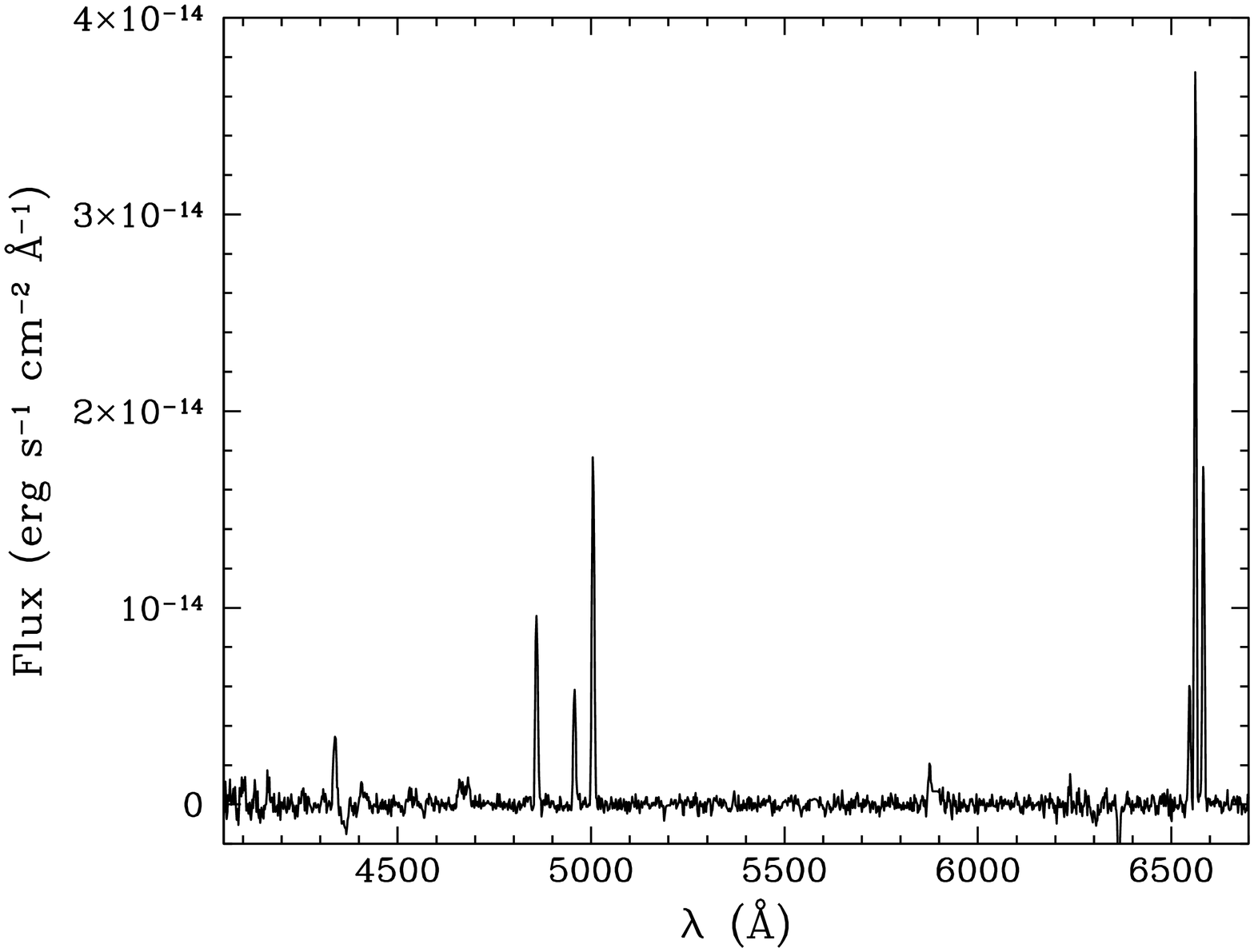} 
\vskip .1in 
\end{center} 
\caption{ 
From top to bottom, SPM B\&C low-dispersion long-slit optical spectra 
of NGC\,650-1 for the eastern bipolar lobe (region 3 of slit D) and 
regions of low- (\textit{middle}) and high-excitation (\textit{bottom})
in the halo (see text for details).
The main spectral lines are identified on the top spectrum. 
} 
\label{boller1.img} 
\end{figure}

The arc-shaped feature presents smooth diffuse H$\alpha$ and [O~{\sc iii}] 
(green-blue in Figure~\ref{650.img}) emissions.   
This arc-like feature is symmetric with respect to the bipolar axis 
of the main nebula, with the H$\alpha$ emission dominating the central 
regions of the arc and the [O~{\sc iii}] emission been more intense at 
its outer edges.   
Interestingly, if we drew two lines connecting the edges of this
arc-like feature with the central star, these lines will touch the
tips of the equatorial ring, thus suggesting the arc-like feature may result from
illumination effects \citep{Kwok2010}.  
The centre of this arc-like feature, if we assume it is an arc of 
a circle, would be located $\sim$60\arcsec\ west-northwest of the 
central star. 
The radius of such circle would be 220\arcsec.

As for the [N~{\sc ii}]-bright western blobs, these are two disconnected 
patches of low-excitation diffuse emission located below and above the 
western bipolar lobe along a line orthogonal to the bipolar axis 
(labelled X and Y in Figure~\ref{650.img}).
The southern blob ``Y'' is brighter and more extended than the
northern blob ``X''.
There is another [N~{\sc ii}]-bright knot just at the tip of the
western bipolar lobe (labelled Z in Figure~\ref{650.img}), but it
seems the tip of a bipolar protrusion which is only faintly
detected.

\subsection{IR imaging}\label{ir_sec}

Near- and mid-IR studies of NGC\,650-1 abound in the literature. 
\citet{Ueta2006} has analysed the 24 $\mu$m, 70 $\mu$m, and 160 $\mu$m 
\emph{Spitzer} MIPS observations of NGC\,650-1, focusing on the main 
nebula. 
The bipolar lobes (and its protrusions) and the equatorial ring of 
NGC\,650-1 are clearly revealed, indeed, in the 24 $\mu$m image 
presented in the middle panel of Figure~\ref{ir.img}. 
\citet{Ueta2006} remarked that the emission from the equatorial ring in 
the 24 $\mu$m image actually peaks inside the edges of the emission from 
ionized material. 
The spatial coincidence of the 24 $\mu$m emission with that of the
He~{\sc ii} emission reported by \citet{Balick1987} indicates that
the emission at this location in this band is dominated by the
[Ne~{\sc v}] 24.3$\mu$m and [O~{\sc iv}] 25.9 $\mu$m emission lines
\citep{vanHoof2013,Clayton2014}.

Otherwise, the 24 $\mu$m \emph{Spitzer} MIPS image discloses an arc 
of diffuse emission towards the East-Southeast which is coincident 
with the arc-like feature seen in the optical H$\alpha$ and [O~{\sc iii}]
emission lines. 
This same structure is traced by the \emph{WISE} W4 band image 
(red colour in the bottom panel of Figure~\ref{ir.img}).

\section{Spectroscopy}\label{spec_sec}

As previously shown, the deep IAC80 image has revealed an interesting outer structure around 
the butterfly-shaped main nebular shell of NGC\,650-1.   
To investigate its nature, we have obtained low- and high-dispersion 
long-slit spectra at selected positions to obtain information on its 
excitation conditions and kinematics.

The high-dispersion echelle data probe both the bipolar main nebula 
and the outer structure.   
The kinematics of the main bipolar nebula has been thoroughly discussed 
in the literature \citep{Taylor1979,Sabbadin1981,Recillas1984,Bryce1996}.
Our long-slit echelle spectra (Figure~\ref{p1.img}) and the resulting 
{\sc SHAPE} modelling are consistent with the distinguished work presented 
by \citet{Bryce1996}. 
The main nebula can be described as a thick equatorial ring that 
collimates two broad bipolar lobes. 
The inclination of the bipolar axis with the line of sight 
is $\sim$85$^\circ$, with the eastern lobe receding from us. 
The de-projected velocity of the bipolar lobes range between 120 and 140 
km~s$^{-1}$, whilst that of the equatorial ring is $\sim$37 km~s$^{-1}$. 
At a distance of 0.93 kpc, the kinematical age is $\simeq$4760 yrs.  
The nebula has a systemic LSR velocity of $-$22 km~s$^{-1}$ as 
derived from the average radial velocity of the equatorial ring. 
Unless stated otherwise, the radial velocities of the different features 
of NGC\,650-1 will be referred to this systemic velocity.

The features in the outermost regions have rather small velocities 
(Figure~\ref{p2.img}). 
The eastern bow-shock has a velocity of $-$2 km~s$^{-1}$ and the 
material behind it $-$3 km~s$^{-1}$ (northern region registered 
by slit \#3) and $\sim$0 km~s$^{-1}$ (western region registered 
by slit \#1).   
The [N~{\sc ii}]-bright blobs have relatively larger velocities,
in the range of $-$10 km~s$^{-1}$.

One-dimensional low-dispersion spectra from 7 different regions have
been extracted to probe the bow-shock structure and regions immediately
behind it.
As marked in the top panel of Figure~\ref{ir.img}, these are the apertures \#1
of slit A, \#1, \#3, and \#6 of slit B, \#1 and \#3 of slit C, and \#4
of slit D.
Similarly, four one-dimensional low-dispersion spectra have been
extracted to probe the [N~{\sc ii}]-bright blobs, namely apertures
\#2, \#4, and \#5 of slit B, and \#2 of slit C. 

Once we have confirmed that the spectra of each morphological component
share the same spectral properties, we have combined them to increase
their S/N ratio.
The combined spectra are presented in Figure~\ref{boller1.img},
corresponding to the [N~{\sc ii}]-bright blobs (middle panel)
and to the bow-shock structure and regions immediately behind
it (bottom panel).

The H$\alpha$ surface brightness of these two regions is $\cong$ 1$\times$10$^{-17}$
and $\cong$ 4$\times$10$^{-17}$ erg cm s$^{-1}$ arcsec$^{-2}$, respectively. 

To compare their spectral properties with those of the main nebula,
one additional spectrum probing the eastern bipolar lobe has been
extracted from the aperture \#3 of slit D.
This spectrum is presented at the top panel of Figure~\ref{boller1.img}.

\begin{figure} 
\begin{center} 
\includegraphics[width=0.9\columnwidth]{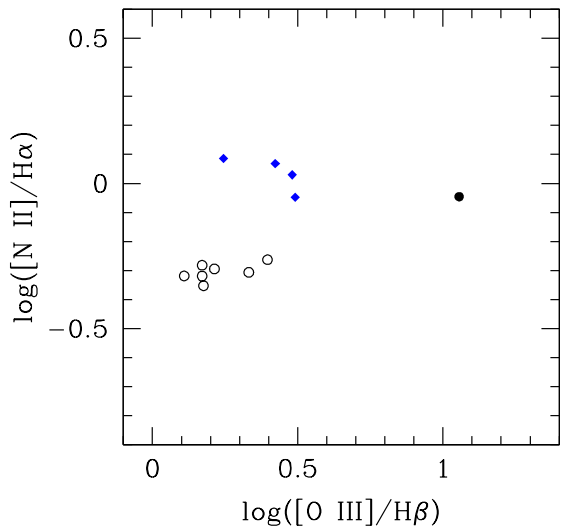} 
\vskip .1in 
\end{center} 
\caption{ 
  Line intensity ratios derived from low-dispersion long-slit optical
  spectra of the bipolar lobes of NGC\,650-1 (black solid dots), the
  outer [N~{\sc ii}]-bright blobs (blue solid diamonds), and the outer
  arc-like features (black open circles).
  }
\label{ratios1.img} 
\end{figure}

We have measured the intensities of the H$\beta$, H$\alpha$, [O~{\sc iii}] 
$\lambda$5007, and [N~~{\sc ii}] $\lambda$6584 in these spectra. 
Their ratios are shown in Figure~\ref{ratios1.img}.   
The low-dispersion spectroscopic data confirm the excitation 
variations through the different components of the outer 
emission seen in the IAC80 image (Figure~\ref{650.img}). 
The [N~{\sc ii}]-bright blobs have [N~{\sc ii}]/H$\alpha$ similar to 
those of the bipolar lobes, but much lower [O~{\sc iii}]/H$\beta$, 
thus implying significantly higher [N~{\sc ii}]/[O~{\sc iii}],
typical for low-ionization structures in PNe \citep{Akras2016}. 
Meanwhile, the bow-shock structure has both lower [N~{\sc ii}]/H$\alpha$ 
and [O~{\sc iii}]/H$\beta$ line ratios than the bipolar lobes, revealing 
the relative prevalence of the recombination lines over the collisionally 
excited lines in this region.

\section{Discussion}\label{sec_dis}

The spatio-kinematical analysis presented in the previous sections implies 
two different structural components in NGC\,650-1: the well-known bipolar 
main nebula and a previously unreported outer structure consisting of an 
arc of smooth emission and several low-excitation blobs.   
We describe next these two structural components.

\subsection{The Main Nebula of NGC\,650-1}

The spatio-kinematical model of the main nebula of NGC\,650-1 presented 
in \S\ref{spec_sec} is very similar to that presented by \citet{Bryce1996}. 
The equatorial region can be described as a thick ring from 
which two bipolar lobes protrude. 
These bipolar lobes show notable extensions at their tips. 
The inclination angle of the equatorial ring and bipolar lobes, in the 
range between 75$^\circ$ and 85$^\circ$, places the nebula symmetry axis 
close to the plane of the sky.

The bipolar lobes of NGC\,650-1 have been described profusely in 
the literature \citep{Sabbadin1981,Recillas1984,Bryce1996,Ueta2006} 
and they will not be discussed here.
As for the equatorial ring, the new images reveal a wealth of complexity. 

There is a significant ionization structure, with the He~{\sc ii} and 
[Ne~{\sc v}] emissions inside the ring, and the [N~{\sc ii}] emission 
delineating its outer edge. 
This betrays the high temperature of its central star, which has been 
estimated to be 208,000 K \citep{vanHoof2013}.
It is thus perplexing that this high-excitation gas inside the equatorial 
ring coexists with dust and H$_2$ molecular material as revealed by the 
presence of dark knots and filaments in the optical images and clumpy H$_2$ 
emission in near-IR images. 
 
This suggests that high density clumps and filaments still 
withstand the action of the stellar wind and high ionization 
stellar flux, although the alignment along the line of sight 
of different gas phases at the ionization front cannot be 
neglected \citep{Ueta2006}.

\subsection{The Halo around NGC\,650-1}

The deep IAC80 image of NGC\,650-1 has revealed a new 
set of features beyond the extent of the main nebula 
(Figure~\ref{650.img}).
This can be classified as a halo ejected during the AGB phase
\citep{Chu1987,Stanghellini1995}.  
The leading edge of the halo, with its bow-shock morphology, is clearly
seen in the \emph{Spitzer} MIPS 24$\mu$m image, but it is undetected in
the 70 $\mu$m and 160 $\mu$m MIPS images \citep{Ueta2006}. 
It is well detected in the \emph{WISE} W4 band 
(red colour in the bottom panel of Figure~\ref{ir.img}).   
Although the emission in the \emph{Spitzer} MIPS 24$\mu$m band of
the equatorial ring of NGC\,650-1 was attributed to the [O~{\sc iv}]
25.9$\mu$ and [Ne~{\sc v}] 24.3$\mu$m line emission, the origin of
this emission in this region of the halo may be different. 
This band also includes dust continuum emission \citep{Chu2009}, which is 
very likely dominant in this region, given its distance to the ionizing 
source. 
This would be consistent with the lower excitation of this region 
revealed by the optical spectra.

The overall morphology of the halo is very reminiscent of that
resulting from the interaction of a low density nebula with the
ISM \citep{Villaver2012,Wareing2007}. 
Therefore, we propose that the outer emission of NGC\,650-1 corresponds
to a halo ejected during the AGB phase which is interacting with the ISM. 
As the star moves through the ISM, the leading edge of the halo is 
compressed, forming the eastern arc-like feature (a bow-shock). 
Meanwhile, material stripped from the halo in its interaction 
with the ISM is seen in the trailing side as blobs and clumps 
particularly bright in the low-excitation line of [N~{\sc ii}].  
The density of these clumps is low, however, as revealed 
by their low H$\alpha$ surface brightness.

This conclusion is supported by its kinematic properties.   
The radial velocity of the different features of the halo registered 
by our long-slit echelle spectroscopy shows small differences, in the 
range between 3 and 10 km~s$^{-1}$, with the systemic velocity of the 
nebula.   
In all cases, the velocity of the halo are blue-shifted with 
respect to the nebula systemic velocity, indicating that the 
halo approaches us, i.e., it is placed between the nebula and 
us. 
If we adopt an expansion velocity of 10 km~s$^{-1}$, then at the 
distance of 0.93 kpc, the kinematical age of this halo would be 
$\sim$80,000 yrs\footnote{ 
  We note the strong dynamical interaction of this halo with the ISM,
  which will affect both its radius and expansion velocity. 
  \citet{2002ApJ...581.1204V} have modelled this
  interaction and concluded that kinematic ages derived for PN
  haloes are unreliable. 
  The kinematic age provided here for the halo of NGC\,650-1 should
  be used as a first order estimate of its age.
}.

Naively, the halo bow-shock structure would be expected to be oriented 
along the direction of the relative motion of the central star of NGC\,650-1 
and the local ISM. 
Under this assumption, the fact that the bow-shock shares its symmetry
axis with the bipolar axis of the main nebula of NGC\,650-1 should be
coincidental, although illumination effects cannot be neglected:
we might just see the section of the leading halo which is  
illuminated by the central star,
resulting in the apparent alignment between the bow-shock at the halo and
the bipolar symmetry axis.   
On the other hand, the protrusions emanating from the eastern lobe 
are indicative of compression, which is further supported by its 
rippled morphology. 
This is not the case for the much weaker protrusions emanating from 
the western lobe.

To determine the motion of the central star of NGC\,650-1 on the plane of the
sky, we have used two \emph{HST} images obtained in 1995.59 and 2009.75 (i.e.,
14.16 years apart) to derive a proper motion of 0\farcs0075~yr$^{-1}$ along
PA$\simeq$112$^\circ$.  
At the distance of 0.93 kpc, this proper motion implies a velocity of 
$\sim$34 km~s$^{-1}$ on the plane of the sky.   
The nebula moves almost parallel to the Galactic Plane, at a height   
below it of $\sim$170 pc.

The systemic velocity of the nebula is $-22$ km~s$^{-1}$, which 
we will adopt as the radial velocity of its central star. 
The motion of the central star along the line of sight, 
approaching us, is consistent with the blue-shifted halo.   
More importantly, the motion of the central star on the plane of the sky 
along PA$\simeq$112$^\circ$ is in great agreement with the symmetry axis 
of the bipolar main nebula at 125$^\circ$.   
It can be concluded that the nebula as a whole, main nebula and halo, 
moves towards the East-Southeast direction, compressing the eastern 
section of the halo into a bow-shock feature and the protrusions from 
the eastern bipolar lobe.   
The trailing material of the halo is seen as the [N~{\sc ii}]-bright blobs.

\section{Summary}\label{sec_con}

We report for the first time the discovery of a halo around the
butterfly PN NGC\,650-1.
This demonstrates that the quotation from Balick's paper ``This is an
outstanding example of an object for which a short exposure tells a
very different story from a long one" \citep{Balick1992} perfectly
applies.
Indeed, they did not detect this halo, but proposed the confinement of
the main nebula by an ``invisible low-density medium'' as suggested by
the absence of ionization fronts and the sharp edges of the lobes. 
Our work confirms the validity of their conclusions, implying that the
``invisible low-density medium'' is real.

It is worthwhile to ask oneself whether all bipolar PNe have spherical
haloes.
Some of them have already been reported in the literature, as
the ones around Vy\,1-2 \citep{Akras2015,Ramos2016} and NGC\,6905
\citep{Rubio2015}.
Many other may wait for detection, but they may be very faint (as the
one detected here in NGC\,650-1) or they might have been stripped by
the interaction with the ISM.  
New deep sensitive images are required for many objects with
similar characteristics.

\section*{Acknowledgments}

GRL acknowledges support from Universidad de Guadalajara 
(Apoyo de Estancias Acad\'emicas -- RG/003/2017), CONACyT, CGCI, PRODEP and 
SEP (Mexico). 
MAG also acknowledges support of the grant AYA 2011-29754-C03-02 and 
AYA 2014-57280-P, both co-funded with FEDER funds. 
LS acknowledges support from PAPIIT grant IA-101316.

Based on observations made with the Observatorio Astron\'omico Nacional at the Sierra de San Pedro M\'artir, OAN-SPM, wich is operated by the Instituto de Astronom\'{\i}a of the Universidad Nacional Aut\'onoma de M\' exico. 
Based on observations made with the Nordic Optical Telescope, operated on the island of La Palma jointly by Denmark, Finland, Iceland, Norway, and Sweden, in the Spanish Observatorio del Roque de Los Muchachos of the Instituto de Astrof\'{\i}sica de Canarias. 
The data presented here were obtained [in part] with ALFOSC, which is provided by the Instituto de Astrof\'{\i}sica de Andaluc\'{\i}a (IAA) under a joint agreement with the University of Copenhagen and NOTSA. 
This article is based on observations made with the IAC80 operated on the island of Tenerife by the IAC in the Spanish Observatorio del Teide.
Based on observations made with the Gran Telescopio Canarias (GTC), instaled in the Spanish Observatorio del Roque de los Muchachos of the Instituto de Astrof\'{\i}sica de Canarias, in the island of La Palma.
This publication makes use of data products from the Wide-field Infrared Survey Explorer, which is a joint project of the University of California, Los Angeles, and the Jet Propulsion Laboratory/California Institute of Technology, funded by the National Aeronautics and Space Administration. 
This work is based [in part] on observations made with the Spitzer Space Telescope, which is operated by the Jet Propulsion Laboratory, California Institute of Technology under a contract with NASA. 
{\sc IRAF}, the Image Reduction and Analysis Facility, is distributed 
by the National Optical Astronomy Observatory, which is operated 
by the Association of Universities for Research in Astronomy (AURA) 
under cooperative agreement with the National Science Foundation.
{\sc XVISTA} is an interactive image and spectral reduction
and analysis package developed at the Lick Observatory
and maintained and distributed by Jon Holtzman at the New Mexico
State University at http://ganymede.nmsu.edu/holtz/xvista/.

\end{document}